\documentstyle[12pt,aps,prb]{revtex}

\newcommand{\p}{\Psi_{\uparrow}}
\newcommand{\q}{\Psi_{\downarrow}}
\newcommand{\r}{\Phi_{\uparrow}}
\newcommand{\m}{\Phi_{\downarrow}}
\newcommand{\be}{\begin{equation}}
\newcommand{\ee}{\end{equation}}  
\newcommand{\ep}{\varepsilon}
\begin{document}

\draft

\title{Local Magnetic Impurities in the 2D   Quantum Heisenberg Antiferromagnet}

\author{V.N. Kotov, J. Oitmaa, and O. Sushkov}
\address{ School of Physics, University of New South Wales, Sydney 2052,
Australia}

\maketitle
\begin{abstract}

 The two-dimensional (2D) quantum Heisenberg antiferromagnet 
at zero temperature, 
 with locally frustrating   magnetic
defects, is studied. 
We consider two types of defects - 
 an isolated ferromagnetic bond (FMB) and a quantum impurity spin,
coupled symmetrically to the two sub-lattices.
A technique is developed to study strong defect-environment
interaction. In the case of a  FMB we find that,
contrary to the previous linear spin-wave result, the local magnetization
stays finite even for strong frustration. For an antiferromagnetic
coupling of a quantum impurity spin with its environment, 
we find a local change in the ground state.
All our calculations are compared with numerical results from exact
diagonalization.

\end{abstract}

\narrowtext

\newpage
\section{Introduction}
 The two-dimensional quantum Heisenberg antiferromagnet (HAFM)
has attracted a lot of attention in the last several years,
mainly because of its relevance to the physics of the
high-$T_{c}$ materials \cite{Manousakis}. 
At zero hole doping, these compounds form planes of localized
Cu spins 
and their physics is well described by the
spin-1/2 square lattice HAFM.
 There is strong numerical evidence that this system has 
long-range order (LRO) in its ground
state at $T=0$. Due to quantum fluctuations,
 the  staggered magnetic moment was found to
be reduced to about $61 \%$ of  the classical value \cite{Manousakis}.  
 Doping with excess oxygen, which is crucial for
superconductivity, creates spin-1/2 holes on the
oxygen sites, located in between the Cu sites. The holes
interact magnetically  with their Cu neighbors. This interaction
leads to frustration of the Cu spins and ultimately to destruction
of magnetic LRO, above a critical dopant
concentration.    

One way to describe the doping process is via an  effective one-band
model, like, e.g. the $t-J$ model \cite{Dagotto}, in which the holes
hop in the antiferromagnetic (AFM) environment.

However,  the extreme limit of static holes,
which frustrate the AFM order only locally, is also believed
to have relevance. The basic idea,
put forward by Aharony {\it et al.} \cite{Aharony},
 is that a hole, interacting
with its two neighbors creates an effective {\it ferromagnetic}
interaction between them. The magnetic properties then
are described by a HAFM with a given concentration of such 
ferromagnetic bonds (FMB). 
As a start, it is important to understand the effect that one
isolated FMB has on the AFM environment.
This problem was   studied by Lee and Schlottmann \cite{Lee}
and Aristov and Maleev \cite{Maleev} 
in the linear spin-wave approximation (LSWA). They found that
the  magnetization of the two spins, connected by the bond
decreases dramatically as the strength of the bond increases,
and noted  an instability that occurs in the theory
at the "classical" transition point (the point where a
local triplet formation is favorable without taking quantum 
fluctuation into account). 
 There have been  several extensions of this work,
taking into account finite temperature and finite
concentration of bonds \cite{Korenblit,Ferrer}.
 One can also treat the frustration effects more realistically,
by considering a model of frustration, in which a static hole
interacts with its two neighbor spins. Oitmaa, Betts, and Aydin
\cite{Jaan}
studied this model numerically, via an exact
diagonalization on small clusters, and found a discontinuous
local phase transition for an antiferromagnetic hole-neighbor
interaction, whereas no change in the ground state occurred
for a ferromagnetic coupling. A similar model was studied
by Clarke, Giamarchi, and Shraiman \cite{Giamarchi} in one dimension 
 via bosonization.

In the present work we study the two models with local
frustration, mentioned above - an isolated
FMB in the 2D HAFM, and a quantum defect spin, coupled 
symmetrically to the two sub-lattices. 
Our motivation is twofold. First, the basic works on the
FMB problem \cite{Lee,Maleev}  use  the LSWA,
which is only valid for small FMB strength. In the
original frustration model however, Aharony {\it et al.} 
\cite{Aharony} concluded that the induced FMB coupling
 should be large. Thus, it is important to study the model
for strong coupling, and, along the way, assess the region
of validity of LSWA. Second, the two-dimensional spin defect model,
which provides a realistic frustration mechanism,
has not been studied analytically, as far as we know.
The present work, thus,  extends the previous work of
  Oitmaa, Betts, and Aydin \cite{Jaan} on this problem.

Our analytical approach is specifically designed
to treat the strong coupling regime. The main idea is that since the
perturbations are local, one should first solve the defect problem
separately, and then take into account the defect-AFM environment
interaction. For the FMB problem this amounts to considering
the two-body Green's function of the two spins, connected by the
bond. Analogously, for the spin defect problem, we introduce
a three-body Green's function (of the spin defect
and its two neighbors). All our analytical results are
compared with numerical simulations, based on exact diagonalization
on small clusters. A short account of the main results
of this work appeared previously in Ref.[10]. 
The rest of the paper is organized as follows.
Sec.II describes the basic idea behind our theoretical
approach and, as a first application, we consider
the HAFM with no  defects. In Sec.III we present our 
analytical, as well as numerical solution
of the FMB problem. Sec.IV deals with the spin defect 
problem from the same perspective. Our conclusions are
summarized in Sec.V.

\section{Method description.}

Consider the two-dimensional spin-1/2 Heisenberg model
on a square lattice at zero temperature:

\begin{equation}
  H_{J} = J \sum_{<i,j>}\vec{S}_{i} . \vec{S}_{j}.
\end{equation}
The summation is over nearest neighbors only, and
$J > 0$. We set $J=1$ from now on. It is known \cite{Manousakis} that there is
LRO in the ground state of $H_{J}$ and the  staggered magnetic
moment $m \equiv <S_{i}^{z}> = 0.303$. 

Now assume  we introduce additional 
 interactions (of typical strength $K$) 
between two  nearest neighbor    spins.
Let us call such a configuration a spin defect.
We have in mind, for example, a FMB, connecting two of the spins.
Since such a  defect  is local, it does not affect
significantly the LRO far away from it.
Therefore, in order to treat the strong coupling  regime ($ K\sim J$),
one  must first take into account the interaction
between the spins in the defect and then treat the
defect-AFM environment interaction. 

As a simple illustration of the above idea consider
the HAFM (1) without any additional interactions.
 Let a "defect" consist of only
one spin $\vec{S}_{1}$, and separate (1)
 into an AFM background (spin-wave) Hamiltonian and a defect-neighbor
interaction:

\begin{equation}
H = H_{sw} + S_{1}^{z}\sum_{i=2}^{5}S_{i}^{z} 
+ \left\{ \frac{1}{2} S_{1}^{+} \sum_{i=2}^{5}S_{i}^{-} +\mbox{ h.c.}
\right\},
% \equiv H_{0} + H_{int}
\end{equation}
\begin{equation}
 H_{sw} = \sum_{\bf k} \varepsilon_{\bf k}(\alpha_{\bf k}^{\dagger}
\alpha_{\bf k} + \beta_{\bf k}^{\dagger} \beta_{\bf k} ).
\end{equation}
Here is $\beta_{\bf k}$ and $\alpha_{\bf k}$ are the usual spin-wave
operators with dispersion \cite{Manousakis}: 

\begin{equation}
\varepsilon_{\bf k} = 2\sqrt{1- \gamma_{\bf k}^{2}}, \\\
 \gamma_{\bf k} = \frac{1}{2}(\mbox{cos}(k_{x}) + \mbox{cos}(k_{y})).
\end{equation}

Next, the Ising part of the interaction in (2) is taken into account in the
mean-field approximation, i.e. we replace:
$\sum_{i=2}^{5}S_{i}^{z} \rightarrow  4 m$.
For the spin $\vec{S}_{1}$ it is convenient to use
the fermion representation:

\be
S^{+}_{1} = \p^{\dagger} \q, \ \  
S^{z}_{1} = \frac{1}{2}(\p^{\dagger}\p - \q^{\dagger} \q),
\ee
where $\p^{\dagger}$ and $\q^{\dagger}$ create
(acting on their common vacuum) a $S_{1}^{z}$ component
$1/2$ and $-1/2$ respectively.
In order for the two fermions to represent  a spin-1/2 operator,
they have to satisfy the constraint: \mbox{ $\p^{\dagger}\p + 
\q^{\dagger}\q =1$}, i.e. the physical states have only one fermion.
In this representation the Hamiltonian becomes:

\be
H = H_{sw} + H_{0} + H_{int},
\ee
\be
H_{0} = 2m (\q^{\dagger}\q - \p^{\dagger} \p),
\ee
\begin{equation}
H_{int} 
 =
 \sqrt{\frac{8}{N}} \p^{\dagger}\q \sum_{\bf k}\gamma_{\bf k}
(u_{\bf k} \beta_{\bf k} + v_{\bf k}\alpha_{-\bf k}^{\dagger})
+ \mbox{ h.c.},
\end{equation}
where $N$ is the total number of lattice sites, and 
 $u_{\bf k} = \sqrt{\frac{1}{2} + \frac{1}{\varepsilon_{\bf k}}} $,
  $v_{\bf k} = -\mbox{sign}(\gamma_{\bf k})
\sqrt{-\frac{1}{2} + \frac{1}{\varepsilon_{\bf k}}} $
are the parameters of the Bogoliubov transformation.
Here we have assumed that the spins $\vec{S}_{i}, i=2,..,5$ belong
to the B sub-lattice (spin down).

We define the Green's
 functions for the two fermion and spin-wave species:
\mbox{ $G_{\mu}(t) = -i<T(\Psi_{\mu}(t) \Psi^{\dagger}_{\mu}(0))>$,
$\mu = \uparrow, \downarrow$},  \ 
\mbox{$D({\bf k},t) =  -i<T(\beta_{\bf k} (t) \beta_{\bf k}^{\dagger}(0))>$} (and
 similarly
for $\alpha_{\bf k}$).
 The unperturbed Green's functions ($H_{int}=0$) are:

\be
G_{\uparrow, \downarrow}(\omega) = \frac{1}{\omega \pm 2m + i\delta}, \ \ 
D({\bf k}, \omega) = \frac{1}{\omega - \epsilon_{\bf k} +  i\delta}.
\ee
Next, the term $H_{int}$ is treated in perturbation theory.
The self-energy to one-loop order is given by the diagram in Fig.1a :

\begin{equation}
\Sigma_{\uparrow}(\varepsilon) =
i\frac{8}{N}\sum_{\bf k}\gamma^{2}_{\bf k} u^{2}_{\bf k}
\int D({\bf k},\ep')G_{\downarrow}(\ep - \ep')
\frac{d\ep'}{2\pi} =
 \frac{8}{N} \sum_{\bf k} \frac{ \gamma^{2}_{\bf k} u^{2}
 _{\bf k}}{ \varepsilon - 2m - \varepsilon_{\bf k}}.
\end{equation}
Thus  the spectral function of the $\p$ particles to this order is
\cite{Mahan}:

\be
A_{\uparrow}(\ep) = -\frac{1}{\pi} \mbox{Im}G_{\uparrow}^{ret}(\ep) = 
\delta(\ep + 2m - \Sigma_{\uparrow}(\varepsilon))
\equiv Z \delta(\ep -\ep^{*}),
\ee
where $\ep^{*}$ is the solution of the equation:

\be
 \ep^{*} + 2m - \Sigma_{\uparrow}(\varepsilon^{*}) = 0.
\ee
The renormalization factor $Z$ is determined from (11):

\be
Z = 1 + \frac{\partial\Sigma_{\uparrow}(\ep^{*})}{\partial \ep}.
\ee
The ground state wave function of the $\Psi$ particle - spin-wave 
system is then:

\be
|G> = \sqrt{Z} \p^{\dag}|0> + \sum_{\bf k} A_{\bf k}
\q^{\dag}\beta^{\dag}_{\bf k}|0>,
\ee

\noindent
where $|0>$  is the vacuum for the $\Psi_{\mu}$ and the spin-wave operators:

\be
\Psi_{\mu}|0> = \alpha_{\bf k}|0> = \beta_{\bf k}|0> = 0.
\ee
From now on we  will also adopt the short-hand notation:

\be
 |\mu> \equiv \Psi_{\mu}^{\dag}|0>, \ \   
 |\mu,\beta> \equiv \Psi_{\mu}^{\dag}\beta^{\dag}_{\bf k}|0>, \ \
 |\mu,\alpha> \equiv \Psi_{\mu}^{\dag}\alpha^{\dag}_{\bf k}|0>, \ \
\mu = \uparrow, \downarrow.
\ee
The  wave function (14) is a sum of coherent (first term) and
incoherent part. 
In the coherent part, $\sqrt{Z}$ represents the probability amplitude
of having the particle $\p$ (i.e. the spin $\vec{S}_{1}$   being up) in
 the ground state of $H$.
The incoherent part comes from the admixture of intermediate
states, each of them entering with probability amplitude

\be
A_{\bf k} = \frac{
<\downarrow,\beta|H_{int}|\uparrow>}{\ep -2m  - \ep_{\bf k}}.
\ee
It is easy to see that:

\be
\sum_{\bf k} |A_{\bf k}|^{2} = 1 - Z,
\ee
which reflects the normalization condition for $|G>$.  
Finally, the magnetization is:

\be
<G|S_{1}^{z}|G> = \frac{1}{2} Z
-\frac{1}{2}(1-Z) = 0.27.  
\ee 
 
The agreement between (19)  and the spin-wave theory result
($m=0.3$)
is quite good. There are two ways to improve our calculation.
First, two-loop (see Fig.2b), and higher order corrections can be included.
Let us  note that due to spin conservation there are no 
  vertex corrections to the self-energy (of the type, shown in
Fig.2c). A similar
phenomenon occurs in the $t-J$ model \cite{oleg}. 
Second, we could define clusters of more than one spin
as our "defect".  Although the calculation accuracy is expected to
improve by taking the above two points into account, the leading
order result is quite reliable, and, for the problems considered
in the next two sections, is in good agreement with numerical
simulations.

\section{Ferromagnetic Bond Problem.}

\subsection{Theory.}

 Consider a ferromagnetic bond  of strength K,
connecting the sites 1 and 2:

\begin{equation}
H = H_{J} - K\vec{S_{1}}.\vec{S_{2}},
\end{equation}
where $K > 0$. The second term in (20) frustrates
the LRO in the neighborhood of the defect and therefore 
 leads to local suppression of the magnetization. 

 Following the idea, outlined in Sec.II, we separate
the interactions into a part within the defect (i.e.
between the spins $\vec{S_{1}}$ and $\vec{S_{2}}$) and
 a defect - spin-wave interaction. The Ising
part of the latter is treated in mean-field approximation,
whereas the transverse part ($H_{int}$ below) is our perturbation:

\begin{equation}
H = H_{sw}  + H_{0}  + H_{int}, 
\end{equation}
\be
H_{0} = -(K-1)\vec{S_{1}}.\vec{S_{2}}+ 3m(S_{2}^{z} - S_{1}^{z}), 
\ee
\begin{equation}
  H_{int} = \sqrt{\frac{1}{2N}} \left \{
S_{2}^{+}\sum_{\bf k} \Gamma({\bf k})  e^{i k_{x}}
(u_{\bf k} \alpha_{\bf k}^{\dagger} + v_{\bf k}\beta_{-\bf k}) +
 S_{1}^{+}\sum_{\bf k} \Gamma({\bf k})
(u_{\bf k}\beta_{\bf k} + v_{\bf k} \alpha_{-\bf k}^{\dagger})
+ \mbox{h.c.}
\right \},
\end{equation}
where 

\be
\Gamma({\bf k}) = 2 \mbox{cos}(k_{y}) + \exp{(ik_{x})},
\ee
and $H_{sw}$ is given by (3). In  (22) and (23)
it is assumed that $\vec{S_{1}}$ belongs to sub-lattice A
(spin up) and  $\vec{S_{2}}$ - to sub-lattice B.

We introduce a fermion representation for the spins $\vec{S}_{1}$  
and  $\vec{S}_{2}$, similarly to (5), and call the corresponding fermions
$\Psi_{\uparrow, \downarrow}$ and $\Phi_{\uparrow, \downarrow}$, respectively. 
Thus (22) and (23) should be thought of as expressed in terms of
these fermions. In order to diagonalize $H_{0}$
it is convenient to define the two-particle matrix Green's
function 
$\hat{G}_{\mu \nu}$:

\be
\hat{G}_{\mu \nu}(t) = -i \left(
\begin{array}{cc}
 <T(\p(t) \m(t) \p^{\dag}\m^{\dag})>& <T(\p(t) \m(t)\q^{\dag}\r^{\dag})>\\
<T(\q(t) \r(t)\p^{\dag}\m^{\dag})>&  <T(\q(t) \r(t) \q^{\dag}\r^{\dag})> 
\end{array}
\right).
\ee
As  (25) suggests, 
the diagonal elements (11 and 22) correspond to the
two-particle 
states $|1> \equiv  |\uparrow\downarrow>$ and
 $|2> \equiv  |\downarrow \uparrow>$, respectively, where
the first arrow represents the spin  $\vec{S}_{1}$ and
the second one - the spin  $\vec{S}_{2}$.
The off-diagonal components represent  transitions between
these two states. For future purposes it is convenient to define
also the states: $|3> \equiv  |\uparrow \uparrow>$ and
$|4> \equiv  |\downarrow\downarrow>$.
 The unperturbed  Green's
functions (corresponding to $H_{0}$) are:

\be
G_{11,22}(\omega) = \frac{1}{\omega - (K-1)/4 \pm 3m +i\delta}, \ \
G_{12}(\omega) = G_{21}(\omega)= \frac{1}{\omega + (K-1)/2 + i\delta}.
\ee
Next, we evaluate the self-energies to lowest order  in perturbation
theory  with respect to  $H_{int}$. 
 It is, in fact, more convenient to use the Releigh-Schr\"{o}edinger
perturbation theory, rather than  Feynman's diagrammatic technique.
%Let us define the states: $|SW\alpha> = \alpha^{\dag}_{\bf k}|SW>$
%and  $|SW\beta> = \beta^{\dag}_{\bf k}|SW>$
%$|SW>$ being the spin wave ground state.
%Notation of the form $|SW, 1>$ means a direct product between
%the spin wave and the spin state.
Using notations, similar to (16),
 the expression for $\Sigma_{11}$ is:

\be
\Sigma_{11}(\ep) = \sum_{\bf k}
\frac{|<3,\alpha|H_{int}|1>|^{2}}{\ep +(K-1)/4 - \ep_{\bf k}}+
\frac{|<4,\beta|H_{int}|1>|^{2}}{\ep +(K-1)/4 - \ep_{\bf k}}.
\ee
Here by $H_{int}$ we mean the  interaction Hamiltonian at fixed wave
vector,
i.e. (23) without the $\bf k$ summation.
Evaluating (27), we get:

\begin{equation}
\Sigma_{11}(\varepsilon)= \frac{1}{N}\sum_{\bf k} \frac{
|\Gamma({\bf k})|^{2} u_{\bf k}^{2}}{
\varepsilon +(K-1)/4 - \varepsilon_{\bf k}}.
\end{equation}
Similar calculations for the other two Green's functions give:

\be
\Sigma_{22}(\varepsilon)= \frac{1}{N}\sum_{\bf k} \frac{
|\Gamma({\bf k})|^{2} v_{\bf k}^{2}}{
\varepsilon +(K-1)/4 - \varepsilon_{\bf k}}, \ \ 
\Sigma_{12}(\varepsilon)= \frac{1}{N}\sum_{\bf k} \frac{
\mbox{Re} \left[\Gamma^{2}({\bf k}) e^{ik_{x}} \right]u_{\bf k}v_{\bf k}}{
\varepsilon +(K-1)/4 - \varepsilon_{\bf k}}.
\ee
This is the one-loop result which is sufficient for our purposes.
Let us mention that  higher loop orders of perturbation theory for
$\hat{G}_{\mu \nu}$ are  more complicated than  those for  the one-particle
propagator of  Sec.II, because vertex corrections are allowed.

In order to evaluate various correlation functions, we
proceed similarly to the calculation of the previous section.
Since the Hamiltonian mixes the fermionic states $|1>$ and
$|2>$, the equation for the effective energy level
$\ep^{*}$, corresponding to (12) now has the form:

\be
\left| 
\begin{array}{cc}
 \ep^{*}-(K-1)/4+3m - \Sigma_{11}(\ep^{*})& -(K-1)/2 + \Sigma_{12}(\ep^{*})\\
 -(K-1)/2 + \Sigma_{12}(\ep^{*})&\ep^{*}-(K-1)/4-3m - \Sigma_{22}(\ep^{*})
\end{array} 
\right| =0.
\ee
The correct normalized eigenstate is of the form
$|12> = \mu_{1}|1> + \mu_{2}|2>$, where $(\mu_{1},\mu_{2})$ is an eigenvector
of the matrix in (30), and $\mu_{1}^{2}+\mu_{2}^{2}=1$.

The ground state wave function  can be written as:

\be
|G> = \sqrt{Z}|12> + \ \mbox{incoherent part},
\ee
where the incoherent part represents the contribution of  the
intermediate states $|3>$ and $|4>$. 
The normalization factor $Z$ is defined as:

\be
Z = 1 + \frac{\partial \tilde{\Sigma}(\ep^{*})}{\partial\ep},
\ee
where the self energy $\tilde{\Sigma}$:

\be
\tilde{\Sigma}(\ep) = \mu_{1}^{2} \Sigma_{11}(\ep) 
+ \mu_{2}^{2}\Sigma_{22}(\ep) + 2\mu_{1}\mu_{2} \Sigma_{12}(\ep).
\ee
The average of any spin operator $\hat{O}$ in the state $|G>$ is:

\begin{eqnarray}
<G|\hat{O}|G>& =&  <12|\hat{O}|12>Z   +
\sum_{i=3\alpha,4\beta}\sum_{\bf k}\frac{|<i|H_{int}|12>|^{2}}{[\ep +(K-1)/4
  - \ep_{\bf
k}]^{2}}<i|\hat{O}|i>    \nonumber\\
 & &= <12|\hat{O}|12>Z   +\frac{1-Z}{2}\sum_{i=3,4}<i|\hat{O}|i>.
\end{eqnarray}
Using the above formula, we get for the magnetization and
the longitudinal and transverse spin-spin correlation functions:

\be
M \equiv  <S_{1}^{z}> = 
\frac{\mu_{1}^{2} - \mu_{2}^{2}}{2}Z,
\ee

\be
C_{L}(1,2) \equiv  <S_{1}^{z}S_{2}^{z}>=
-\frac{1}{4} -\frac{1}{2}(Z-1),
\ee

\be
C_{T}(1,2) \equiv 
\frac{1}{2}<S_{1}^{+}S_{2}^{-} + \mbox{h.c.}> = \mu_{1} \mu_{2} Z.
\ee
The numerical evaluation of the sums (28) and (29) and the solution of
(30) are straightforward. The correlators are then computed by using
Eqs.(32,33,35-37). 

\subsection{Exact diagonalization studies.}

To estimate the accuracy of our analytic approach
we have obtained numerical results by exact
diagonalization of small clusters. Instead of using
periodic boundary conditions, we have applied
a staggered magnetic field in the $z$ direction to spins on
the boundary of the cluster. This breaks the sublattice
symmetry and allows us to compute single-spin averages
as well as to distinguish between longitudinal and transverse
correlations.

This procedure of course also breaks the translational invariance
of the cluster and will, in general, lead to increased finite size
effects and slow convergence to the bulk limit. We have chosen a cluster
of N = 18 sites and a boundary field to give $<S^{z}>=0.3$ on
the boundary spins. Extremely good qualitative agreement was obtained 
with the analytical results, and tuning of the field would achieve
even better agreement. This however is not warranted without, at the
same time, a systematic extrapolation of the cluster results
to the thermodynamic limit.

\subsection{Results.} 
The results obtained by the Green's function method
as well as the exact diagonalization results are summarized in Figures 2 and 3.

Figure 2 plots the local magnetization $M$ and
the total spin-spin correlation function across the FMB
$C(1,2) \equiv  < \vec{S}_{1}. \vec{S}_{2}> = C_{L}(1,2)+ C_{T}(1,2)$
  as a function of
frustration $K$. We have also calculated $M$ by using LSWA,
essentially following Lee and Schlottmann \cite{Lee}.
By treating $K$ as a perturbation one can, in fact, exactly solve
the Dyson equation for the spin-wave Green's function \cite{Lee}. The 
LSWA result shows
that the magnetization vanishes around $K=1.9$
and an instability occurs at $K=2$, which is the 
point when a local triplet formation is expected
to occur classically.

Our result (35) shows qualitatively different behavior.
Up to $K=1$ the two curves follow each other closely, both
predicting a slight increase of $M$ at that value
(corresponding to a missing bond). A similar suppression of
quantum fluctuations was observed in the case of a missing site
\cite{Bulut}. However, beyond $K=1$ we predict that
the local magnetization decreases slowly and stays non-zero
even for large $K$. On the other hand, the correlator $C(1,2)$
 changes sign
and becomes ferromagnetic at around $K=2.1$. 
As shown in Figure 3, it is, in fact, the transverse part of
the correlator that changes sign, while the Ising part
has antiferromagnetic sign for all $K$.
With increasing $K$ the transverse part increases and the Ising
one decreases (in magnitude), but remains non-zero, which is
consistent with a finite magnetization.

The exact diagonalization results    agree qualitatively well with
those obtained by the Green's function method.
 
We would like to stress that, as our results show,
the range of validity of the LSWA is limited to small 
frustration only ($K<1$). Beyond this  point the number
of generated spin waves is large and the interactions
between them become important. The vanishing of 
the magnetization in LSWA is thus an artifact of the  approximation.
Our treatment, on the other hand, is applicable
far beyond  $K=1$.

\section{Impurity spin problem.}

In this section we consider local frustration
due to an additional quantum spin \mbox{ $\vec{\sigma}$ ($\sigma = 1/2$)}, coupled
symmetrically to the two sub-lattices: 

\be
H = H_{J} + L\vec{\sigma} .(\vec{S_{1}} + \vec{S_{2}}) 
- \vec{S_{1}}.\vec{S_{2}}.
\ee
For either sign of the coupling $L$, the second term in (38)
frustrates the spins $\vec{S_{1}}$ and $\vec{S_{2}}$.
It is also assumed that the impurity spin effectively
removes the superexchange between them \cite{Jaan}.

Similarly to the approach used in Sec.III, first we  diagonalize
the second term in (38) exactly, and then treat 
 the interaction of  $\vec{S_{1}}$ and $\vec{S_{2}}$ with
the AFM background (spin waves)  perturbatively.
Thus we have:

\be
H = H_{sw} + H_{0} + H_{int},
\ee
where

\be
H_{0} =  L\vec{\sigma} .(\vec{S_{1}} + \vec{S_{2}})
+3m(S_{2}^{z} - S_{1}^{z}),
\ee
and $H_{sw}$ and $H_{int}$ are given by (3) and (23) respectively.
We have assumed, as in the derivation of (23), that
$\vec{S_{1}} \in \mbox{(sub-lattice A)}$,
 $\vec{S_{2}} \in \mbox{(sub-lattice B)}$.

In order to diagonalize $H_{0}$, define the bare states
(the arrows represent, from left to right, the $z$ components of  $\vec{S_{1}}$,
$\vec{\sigma}$ and $\vec{S_{2}}$):

\begin{eqnarray}
|1> = |\uparrow \uparrow \downarrow>,|2> = | \downarrow \uparrow 
\uparrow>,|3> = |\uparrow  \downarrow \uparrow>,& S_{tot}^{z} =1/2, & \\
|4> = |\uparrow\downarrow\downarrow>, |5> = |\downarrow\downarrow
\uparrow>,|6> = | \downarrow \uparrow\downarrow>,& S_{tot}^{z} = -1/2,  & \\
|up> = |\uparrow \uparrow\uparrow >,& S_{tot}^{z} = 3/2.  &
\end{eqnarray}
Observe that  $S_{tot}^{z} = 
\sum_{i}S_{i}^{z} + \sigma^{z}$ is conserved and thus each
sector can be dealt with separately.
Diagonalizing $H_{0}$ in the sectors  $S_{tot}^{z}=\pm1/2$:
\be
\mbox{det}(<i|H_{0}|j> - \ep \delta_{ij})=0 
\ee
we get the  eigenenergies $\ep_{i}$ which correspond
to the eigenvectors: 
\be
|i+> = \sum_{j=1}^{3} a_{ij}|j>, S_{tot}^{z}=1/2,
\ee
\be
|i-> = \sum_{j=4}^{6} a_{ij}|j>,  S_{tot}^{z}=-1/2.
\ee
Both states are assumed to be normalized.
The above two sectors 
have the lowest energy (compared to $S_{tot}^{z} =\pm 3/2$) and are degenerate.
For definiteness  we choose, for example
 $S_{tot}^{z} =1/2$. It  will be proven below,
both analytically  and numerically that the ground state
of $H$ stays in this sector for any $L$ even after $H_{int}$
is taken into account.

In order to develop a perturbation theory in  $H_{int}$ it
is convenient to introduce  Green's functions.
 Since there are three states of three particles,
we have to consider a $3\times3$ matrix three-particle Green's function
$\hat{G}_{ij}, i,j=1,2,3$. The poles of $\hat{G}_{ij}$ are the matrix elements
$<i|H_{0}|j>$. In order to define $\hat{G}_{ij}$ let us introduce
a fermionic operator representation  for the three spins, analogously
to (5). The fermions  representing $\vec{S}_{1}$, $\vec{S}_{2}$ and
$\vec{\sigma}$  are  $\Psi_{\mu}$,  $\Phi_{\mu}$ 
and $\Theta_{\mu}$, respectively, $\mu = \uparrow, \downarrow$.
 The Hamiltonian now should be thought of
as expressed in terms of these operators. Then, for example  $\hat{G}_{11}$ is defined
as:

\be
\hat{G}_{11}(t) =-i<T(\Psi_{\uparrow}(t)\Theta_{\uparrow}(t)
\Phi_{\downarrow}(t)\Psi^{\dag}_{\uparrow}\Theta^{\dag}_{\uparrow}
\Phi^{\dag}_{\downarrow})>,
\ee
and similarly for all the others.
The unperturbed Green's functions are:
\be
\hat{G}^{-1}_{ij}(\omega) = \omega -  <i|H_{0}|j> + i\delta,
\ \ i,j=1,2,3.
\ee
To lowest order of perturbation theory with respect to $H_{int}$,
the self-energies, corresponding to $\hat{G}_{ij}$ are calculated as:

\begin{eqnarray}
\Sigma_{ij}(\ep)& =& \sum_{l=1}^{3} \sum_{\bf k}
\frac{<i+|H_{int}|l-,\beta><l-,\beta|H_{int}|j+>}
{\ep - \ep_{l} - \ep_{\bf k}}+ \nonumber\\
 & &\sum_{\bf k}\frac{<i+|H_{int}|up,\alpha><up,\alpha|H_{int}
|j+>}{\ep - \ep_{up} - \ep_{\bf k}},
\end{eqnarray}
where $\ep_{up} = L/2$ is  the energy, corresponding
to the state $|up>$ (43).

Then the equation for the effective energy level $\ep^{*}$
(compare with (30)) is:
\be
\mbox{det}((\ep_{i} - \ep^{*})\delta_{ij} + \Sigma_{ij}(\ep^{*}))=0.
\ee
The  correct three-particle $S_{tot}^{z}=1/2$ eigenstate is given by:
\be
|123> = \sum_{i=1}^{3} \mu_{i}|i+>.
\ee
Here $\mu_{i}$ are the normalized eigenvectors of the matrix in (50).
The three-particle - spin-wave ground state to lowest order
is, similarly to (31):

\be
|G> = \sqrt{Z}|123> + \mbox{incoherent part}. 
\ee
The normalization factor $Z$ is defined as:

\be
Z = 1 + \sum_{i,j=1}^{3} \mu_{i}\mu_{j} \frac{\partial \Sigma_{ij}(\ep^{*})}
{\partial \ep}.
\ee
The incoherent part is a mixture of the intermediate
states $|i-,\beta>$ and $|up,\alpha>$ with appropriate weights, which are
  too
lengthy to write  down explicitly.

Finally, the average of any combination of spin operators $\hat{O}$
in this ground state
can be easily calculated:

\begin{eqnarray}
<G|\hat{O}|G> = <123|\hat{O}|123>Z + \sum_{\bf k}
\frac{|<up,\alpha|H_{int}|123>|^{2}} 
{(\ep - \ep_{up} - \ep_{\bf k})^{2}} <up|\hat{O}|up> + \nonumber \\
\sum_{l,m=1}^{3} \sum_{\bf k} \frac{<123|H_{int}|m-,\beta>
<l-,\beta|H_{int}|123>}{(\ep - \ep_{m}- \ep_{\bf k})
(\ep - \ep_{l}- \ep_{\bf k})}<m-|\hat{O}|l->.
\end{eqnarray}
Using (54) we have calculated the spin-spin correlation
functions for $\vec{S}_{1,2}$ and $\vec{\sigma}$, as well as
the appropriate magnetizations.
The following notations are adopted:

\be
C(\sigma,i) \equiv <\vec{\sigma}.\vec{S}_{i}>, i= 1,2,
\ \ C(1,2)  \equiv <\vec{S}_{1}.\vec{S}_{2}>, 
\ee
\be
M(\sigma) \equiv <\sigma^{z}>, \  M(i) \equiv <S_{i}^{z}>, i= 1,2
\ee

We have also performed numerical simulations, based on exact
diagonalization on small clusters with typical
size $N=18+1$. The numerical procedure
is similar to the one outlined in Sec.IIIB.
 All our results are summarized in Figures 4 and 5.
We have checked with  both methods  that
the ground state is in the sector $S_{tot}^{z}=1/2$ for
all $L$ and thus no level crossing occurs.

 For ferromagnetic value of the coupling, $L<0$,
all correlations become ferromagnetic for sufficiently
large $|L|$(Fig.4.). However, this is not accompanied
by a change in the ground state, since all the three spins
have a non-zero magnetization (Fig.5.). This was already
pointed out by Oitmaa, Betts, and Aydin \cite{Jaan}.
The situation is rather similar to the case  of a FMB in an
AFM background, studied in the previous section.

 For antiferromagnetic  coupling, $L>0$, there is
a local ground state  phase transition at $L\approx 2.3$,
when the spins $\vec{S}_{2}$ and $\vec{\sigma}$ change 
the direction of their magnetization. Naturally,
the magnetization of $\vec{S}_{1}$  increases at
this point, since it is no longer frustrated.
 This local change in the ground state was 
observed numerically by Oitmaa, Betts, and Aydin \cite{Jaan}.
However, they found a discontinuity in all correlation
functions as well as the magnetizations at the transition point, 
which was attributed
to  level crossing. We, on the other hand, observe
a continuous transition. It is quite possible
that the discontinuity is due to a finite size effect.

\section{Summary and conclusions.}

 In summary, we have developed an analytic technique
which allowed as to study the effect of locally frustrating
 perturbations (static spin defects)
in 2D quantum antiferromagnets  for any strength of the coupling.
 Such defects are of relevance to  the physical system of
oxygen holes in the CuO planes of the high-$T_{c}$ cuprates at low doping.
All analytic results are in good agreement with numerical simulations,
based on exact diagonalization on small clusters.
 
 For an isolated ferromagnetic bond we found that the local
magnetization does not vanish even for strong frustration.
 Our result suggests, that the previous calculations using
LSWA \cite{Lee,Maleev}  lead to qualitatively wrong behavior.
The region of validity of LSWA is thus restricted to
small frustration only.

 We also studied frustration due to a quantum impurity spin
coupled symmetrically to the two sub-lattices. 
It was found that for a ferromagnetic sign of the coupling 
there is no change in the ground state, similarly to
the case of a FMB. For an antiferromagnetic interaction between
the impurity and its neighbors we report a local change in the
ground state.

The technique presented in this paper is not limited 
to the study of the above problems.  It can be applied
to any spin systems where  LRO is suppressed locally
due to additional interactions  and represents a 
nice alternative to LSWA for strong interactions.

\acknowledgments
The financial support of the Australian Research Council is
gratefully acknowledged.

%\appendix
%\section{Bla}

\vspace{4cm}

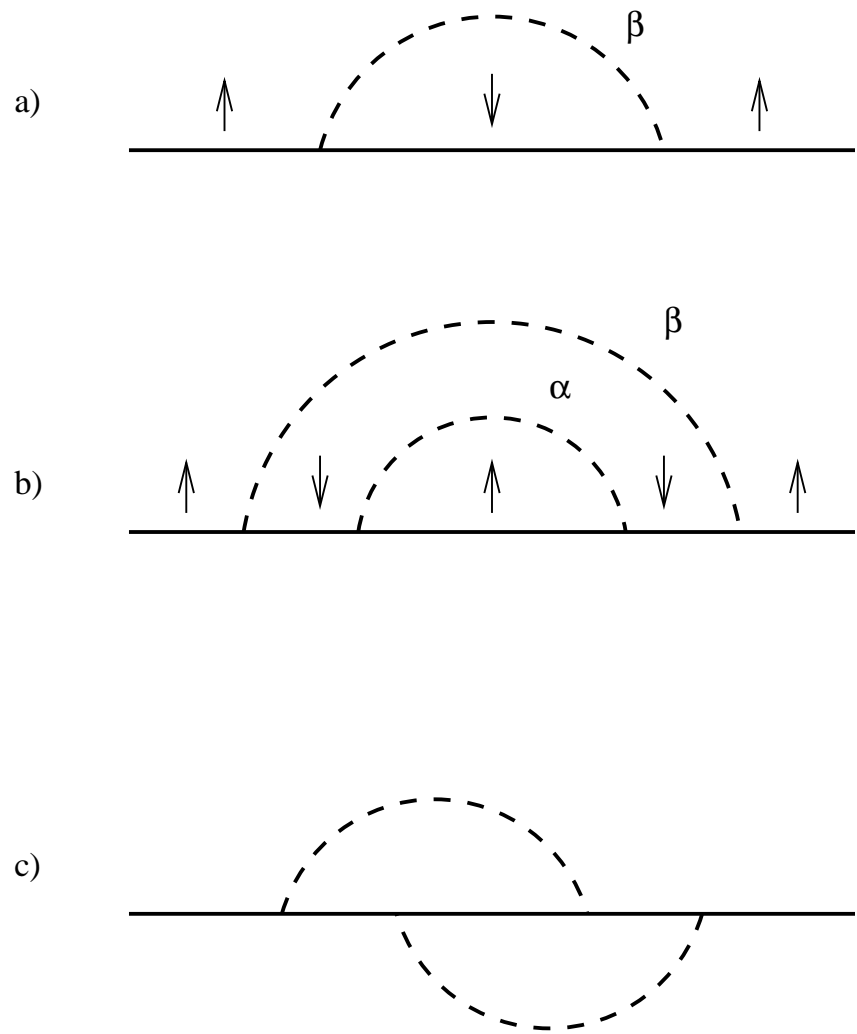
\begin{figure}
\caption{Diagrams for the self-energy $\Sigma_{\uparrow}$:
a) One-loop contribution. b),c) Examples of two-loop corrections.
Solid lines  represent   fermionic Green's functions,
while  dashed lines are  spin-wave propagators.}
\label{fig1}
\end{figure}

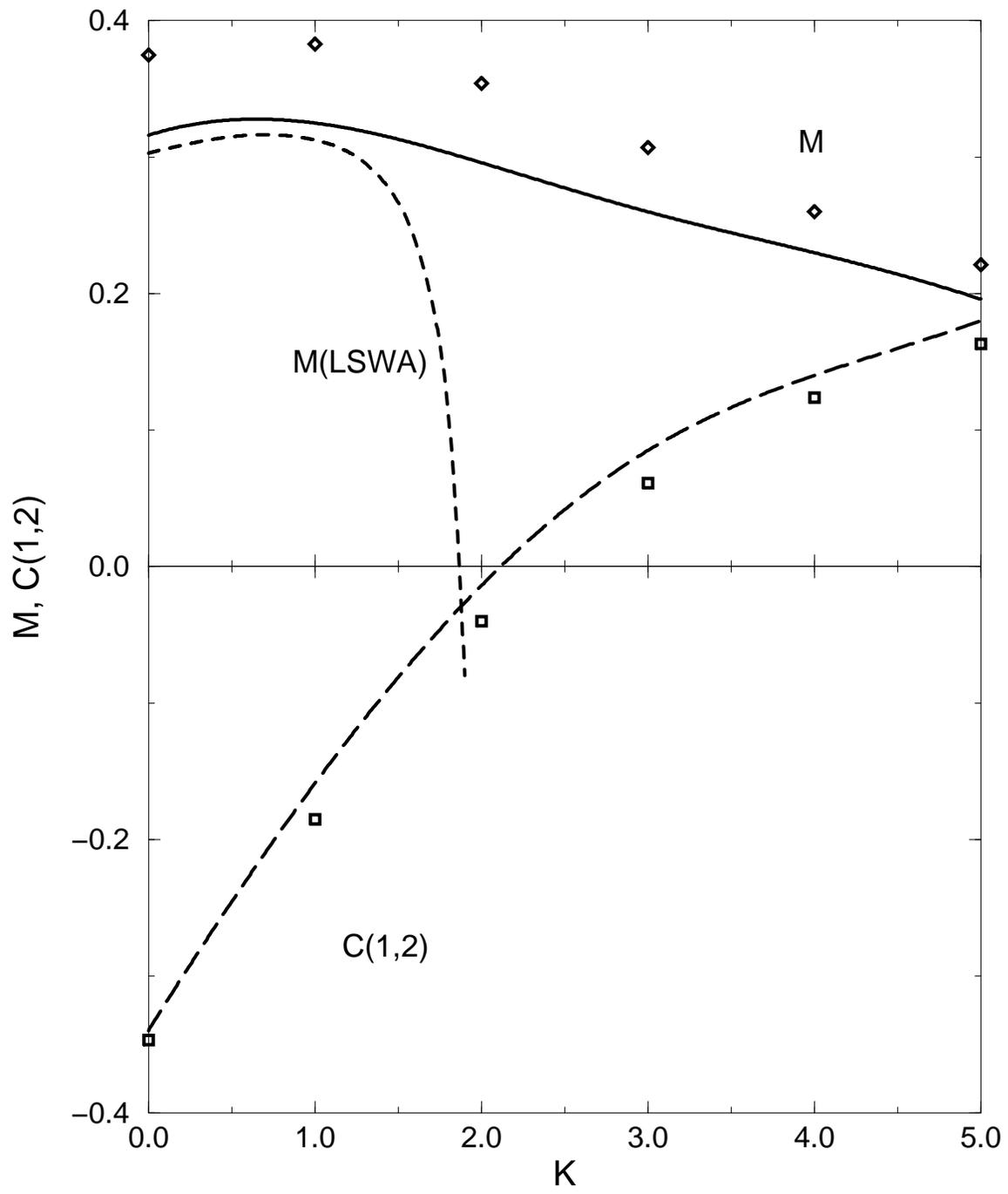
\begin{figure}
\caption{
 Local magnetization $M$ (solid line) and the correlation
function $C(1,2)$ (long dashed line), calculated by using 
the Green's function method, as a function of the FMB strength $K$.
The diamonds and squares are the corresponding exact diagonalization results.
The short dashed line is the magnetization, calculated in LSWA. 
}
\label{fig2}
\end{figure}

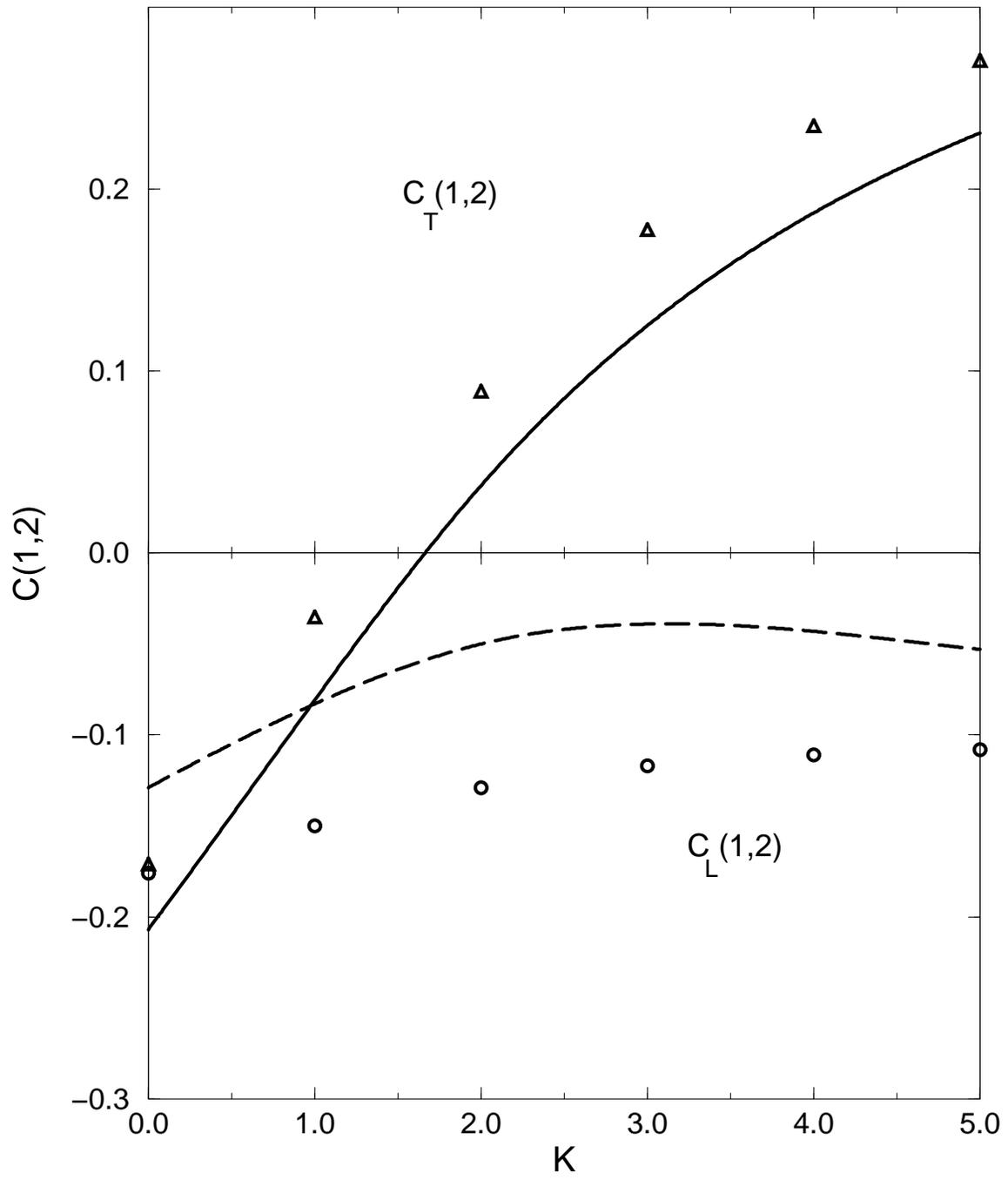
\begin{figure}
\caption{
 The transverse (solid line) and the longitudinal (dashed line)
parts of the spin-spin correlation function across the FMB.
 The triangles and the circles
are the corresponding exact diagonalization results.
}
\label{fig3}
\end{figure}

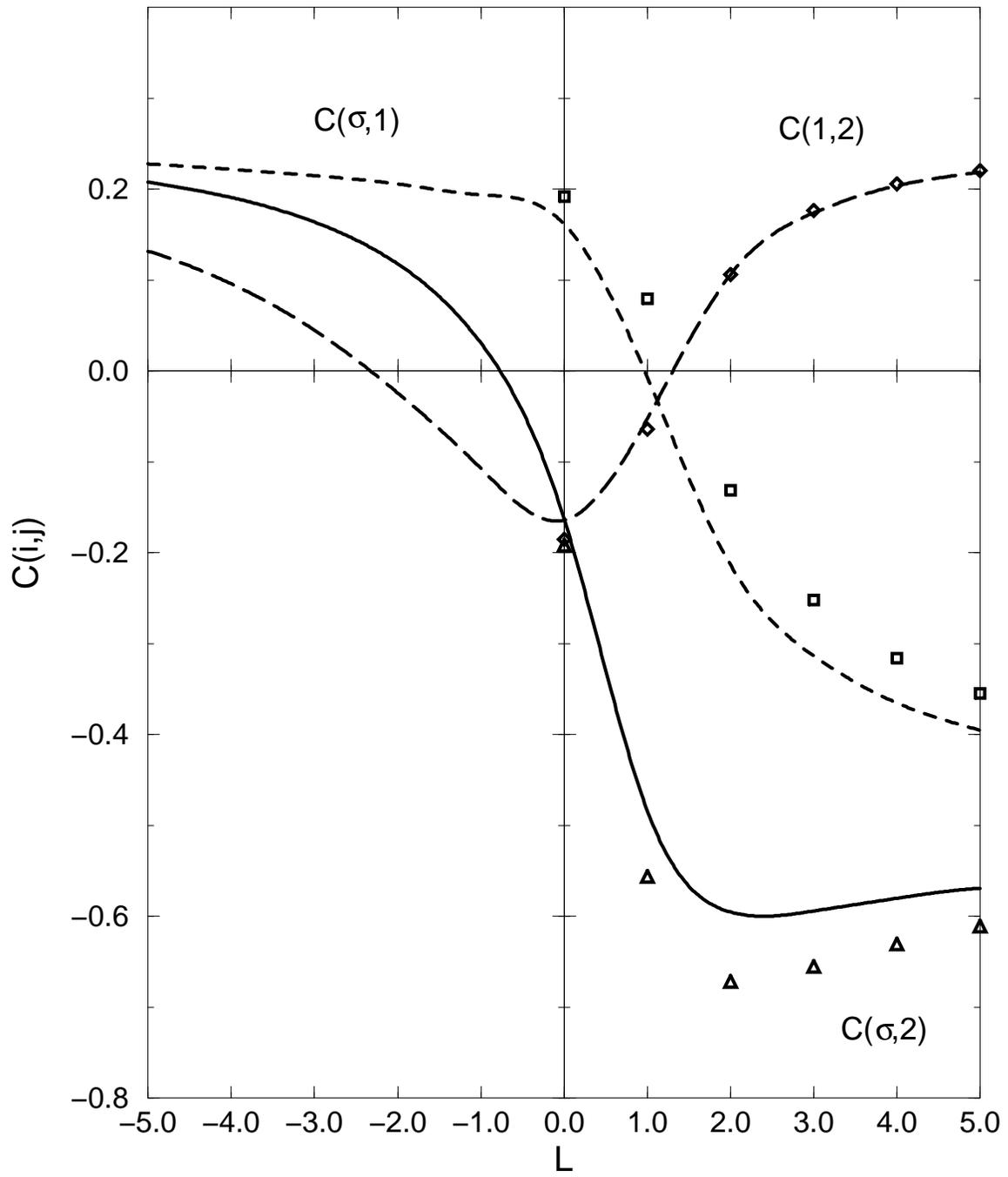
\begin{figure}
\caption{
 Correlations between the impurity spin and its neighbors, as defined
in Eq.(55). The
lines and the symbols represent, respectively, the analytical and the
 exact diagonalization  
results.}
\label{fig4}
\end{figure} 

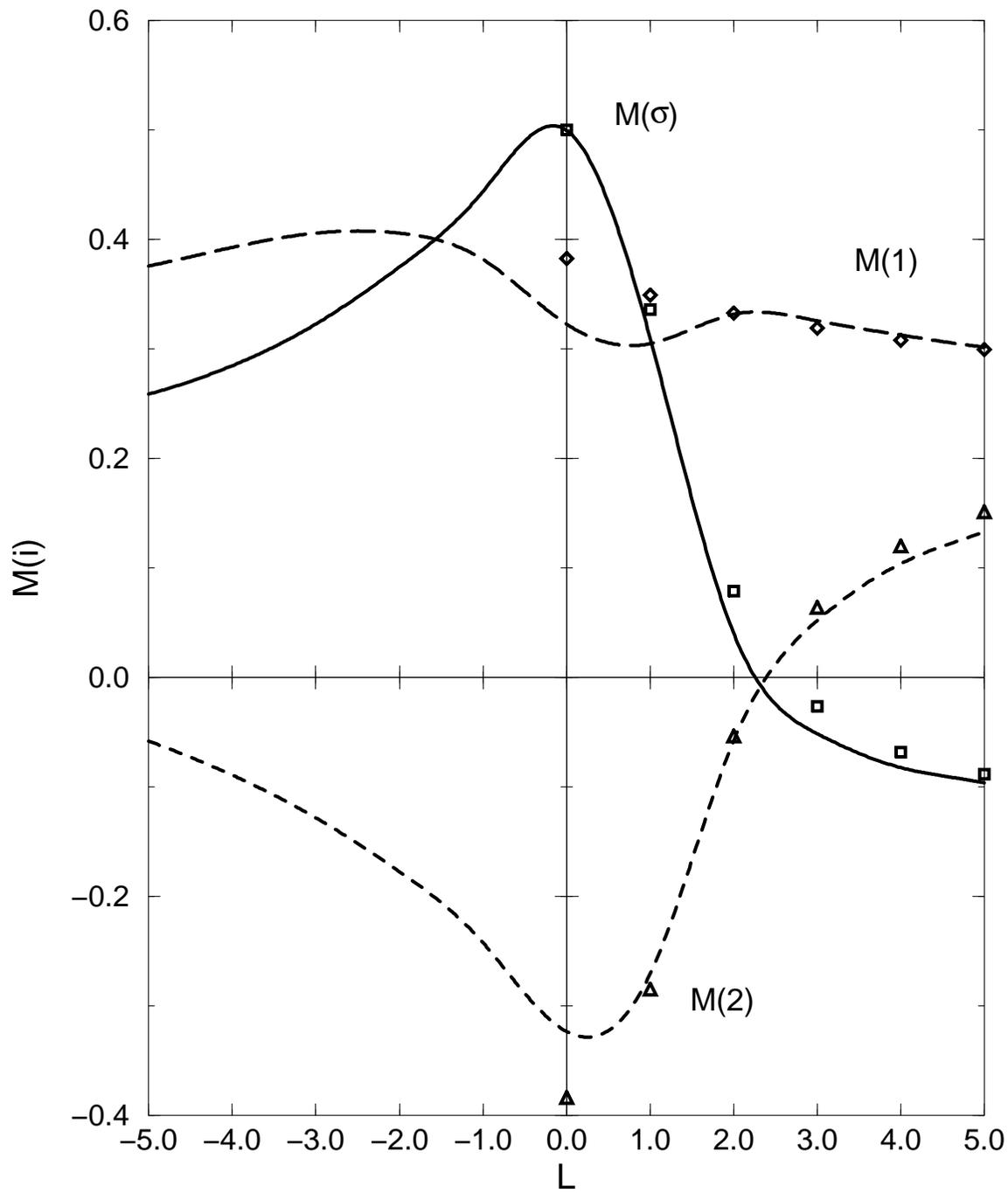
\begin{figure}
\caption{
 The magnetization of the impurity (solid line) and the
neighboring spins (dashed lines) as a function of the interaction $L$.
The symbols represent the exact diagonalization results.
}
\label{fig5}
\end{figure}


\begin{references}

\bibitem{Manousakis} For a review, see, E. Manousakis, Rev. Mod. Phys. {\bf 63}, 1
 (1991).

\bibitem{Dagotto} See, e.g.,
E. Dagotto, Rev. Mod. Phys. {\bf 66}, 763 (1994).


\bibitem{Aharony}A. Aharony, R. J. Birgineau, A. Coniglio, M. A. Kastner,
and H. E. Stanley, Phys. Rev. Lett.  {\bf 60}, 1330 (1988). 

\bibitem{Lee} K. Lee and P. Schlottmann, Phys. Rev. B {\bf 42}, 4226 (1990);
S. Haas, D. Duffy, and P. Schlottmann,  Phys. Rev. B {\bf 46}, 3135  (1992).

\bibitem{Maleev} D. N. Aristov and S. V. Maleev, Z. Phys. B {\bf 81},
433 (1990).

\bibitem{Korenblit} I. Ya. Korenblit, Phys. Rev. B {\bf 51}, 12551 (1995).


\bibitem{Ferrer} J. R. Rodriguez, J. Bonca, and J. Ferrer,
Phys. Rev. B {\bf 51}, 3616 (1995). 


\bibitem{Jaan}  J. Oitmaa and D. D. Betts, Physica A {\bf 177}, 509 (1991); 
 J. Oitmaa, D. D. Betts, and M. Aydin, Phys. Rev. B {\bf 51},
2896 (1995).


\bibitem{Giamarchi} D. G. Clarke, T. Giamarchi, and B. I. Shraiman,
Phys. Rev. B {\bf 48}, 7070 (1993).



\bibitem{us} V. Kotov, J. Oitmaa and O. Sushkov,
J. Magn. Magn. Mat. {\bf 177}-{\bf 180}, (1998).

\bibitem{Mahan} G. D.  Mahan, {\it Many Particle Physics},
Plenum, New York, 1990.

\bibitem{oleg} O. P. Sushkov,  Phys. Rev. B {\bf 49}, 1250 (1994).

\bibitem{Bulut} N. Bulut, D. Hone, D. J. Scalapino, and
E. Y. Loh, Phys. Rev. Lett. {\bf 62}, 2192 (1989).

\end{references}
\end{document}